\documentclass[12pt]{article}

\begin{document}

\title{\bf Deformations of the Boson $sp(4,R)$ Representation and its Subalgebras}
\author{J.P.Draayer$^{1}$, A.I.Georgieva$^{2}$ and M.I.Ivanov$^{2}$
\\
\\
$^{1}${\it Louisiana State University,}\\
{\it Department of Physics and Astronomy,}\\
{\it \ Baton Rouge, Louisiana, 70803USA}\\
\\
$^{2}${\it Institute of Nuclear Research and Nuclear Energy,}\\
{\it \ Bulgarian Academy of Sciences,}\\ 
{\it Sofia 1784, Bulgaria}}

\maketitle
\vspace{.15cm}

\begin{abstract}
\tiny{
The boson representation of the $sp(4,R)$ algebra and two distinct
deformations of it, $sp_{q}(4,R)$ and $sp_{t}(4,R)$, are considered, as well
as the compact and noncompact subalgebras of each. The initial as well as
the deformed representations act in the same Fock space, $\mathcal{H}$,
which is reducible into two irreducible representations acting in the
subspaces $\mathcal{H}_{+}$ and $\mathcal{H}_{-}$ of $\mathcal{H}$.

The deformed representation of $sp_{q}(4,R)$ is based on the standard
$q$-deformation of the boson creation and annihilation operators. The subalgebras
of $sp(4,R)$ (compact $u(2)$ and noncompact $u^{\varepsilon }(1,1)$ with
$\varepsilon =0,\pm $) are also deformed and their deformed representations are
contained in $sp_{q}(4,R)$. They are reducible in the $
\mathcal{H}_{+}$ and $\mathcal{H}_{-}$ spaces and decompose into irreducible
representations. In this way a full description of the irreducible unitary
representations of $u_{q}(2)$ of the deformed ladder series $u_{q}^{0}(1,1)$ and
of two deformed discrete series $u_{q}^{\pm }(1,1)$ are obtained.

The other deformed representation, $sp_{t}(4,R)$, is realized by means of a
transformation of the $q$-deformed bosons into $q$-tensors (spinor-like)
with respect to the $su_{q}(2)$ operators. All of its generators are
deformed and have expressions in terms of tensor products of spinor-like
operators. In this case, a deformed $su_{t}(2)$ appears in a natural way as
a subalgebra and can be interpreted as a deformation of the angular momentum
algebra $so(3)$. Its representation in $\mathcal{H}$ is reducible and
decomposes into irreducible ones that yields a complete description of the
same.

The basis states in $\mathcal{H}_{+}$, which require two quantum labels, are
expressed in terms of three of the generators of the $sp(4,R)$ algebra and
are labeled by three linked integer parameters.
}
\end{abstract}

\section{Introduction}

Symplectic algebras enter in physical applications when operators that
change the number of particles of the system are employed. One example is a
description of collective vibrational excitations of a system of particles
moving in an $n$-dimensional harmonic oscillator potential. In this paper we
start by considering the simplest two-dimensional case with $sp(4,R)$ as its
dynamical symmetry algebra. $Sp(4,R)$ is a noncompact group that is
isomorphic to $O(3,2)$ \cite{Goli}. A reduction from $sp(4,R)$ to the $
u(2)=su(2)\oplus u(1)\sim so(3)\oplus o(2)$ subalgebra gives rise to a
classification scheme with basis states that exhibit collective rotations.
There are three possible reductions to representations of the noncompact $
u(1,1)=su(1,1)\oplus u(1)$ subalgebra that are important for a complete
classification of the basis states of a system.

Although $sp(4,R)$ is the simplest nontrivial case of a noncompact
symplectic algebra, this structure is realized in various applications \cite
{Roro,pehe}. It also serves as an example for exactly solvable test models 
\cite{Goli}. The applications are related to different interpretations of
the quantum numbers of the bosons used to create its representations. In
addition to its use as the dynamical symmetry in some collective models of
nuclear structure \cite{clas}, $sp(4,R)$ has been used for a complete
classification of yrast-band energies in even-even nuclei \cite{clus}. And
since it is rather easy to generalize $sp(4,R)$ results to higher rank
algebras \cite{clas}, features uncovered for $sp(4,R)$ have extended
applications, the bosonisation of other symplectic algebras being a case in
point \cite{qmo}. A further application of $sp(4,R)$ is in the application
of mapping methods \cite{klmap}, where the main purpose is to simplify the
Hamiltonian of the initial problem \cite{asgf}. In all such applications, a
tensor realization of the $sp(4,R)$ algebra derived from the usual boson
creation and annihilation operators, is most convenient.

In the last decade a lot of effort, from a purely mathematical as well as a
physical point of view \cite{bid},\cite{Hay} has been concentrated on
various deformations of the classical Lie algebras. The general feature of
these deformations is that at some limit of the deformation parameter $q$,
the $q$-algebra reverts back to a classical Lie algebra. More than one
deformation can be realized for one and the same ``classical'' algebra,
which can be exploited in different physical applications. There are a lot
of similarities between the classical Lie algebras and their deformations,
especially with respect to the action spaces of their representations. The
study of deformed algebras can also lead to deeper understanding of the
physical significance of the deformation.

In this paper we explore boson representations of the $sp(4,R)$ algebra. We
begin with the well-known representation of this algebra in terms of
``classical'' boson creation and annihilation operators ( Section 1) and
consider all the subalgebras and various ways to specify basis states by
means of eigenvalues of the operators associated with them. We also
introduce a deformation of this algebra in terms of standard $q$-bosons, and
following the same procedure we investigate the enveloping algebra of $
sp(4,R)$ that is so obtained and explore the action of its generators on the
basis, which remains the same (Section 2). We obtain another deformation of
the same algebra by transforming the $q$-deformed bosons into tensor
operators with respect to the compact subgroup $SU_{q}(2)$ defined in the
previous section. In this case we use $q$-tensor products to obtain its
generators, which are also tensor operators in respect to $SU_{q}(2)$. Their
components form subalgebras in a natural way and the compact subalgebra $
su_{t}(2)$ that is so obtained can be interpreted as isomorphic to a
deformation of the $so(3)$ algebra (Section 3). In the last section (Section
4) we investigate a representation of the basis in terms of generators of
the $sp_{t}(4,R)$ algebra, which introduces three quantum numbers for
specifying states that map onto the corresponding classical results.

\section{Boson representations of sp(4,R) algebra.}

Let us begin by recalling some features of the boson representation of $
sp(4,R)$ \cite{bar,clas}. The operators $b_{i}^{\dagger }$ and $b_{i},i=\pm
1,$ which satisfy Bose commutation relations 
\begin{equation}
\begin{array}{ll}
\lbrack b_{i},b_{k}^{\dagger }]=\delta _{i,k}, & [b_{i}^{\dagger
},b_{k}^{\dagger }]=[b_{i},b_{k}]=0 ,
\end{array}
\label{cc}
\end{equation}
are the natural language for a description of the two-dimensional harmonic
oscillator \cite{Goli}. They act in a Hilbert space $\mathcal{H}$ with a
vacuum $|0\rangle$ so that $b_{i}$ $|0\rangle =0.$ The scalar product in $
\mathcal{H}$ is chosen so that $b_{i}^{\dagger }$ is the Hermitian conjugate
of $b_{i}$ $[(b_{i}^{\dagger })^{*}=b_{i}]$ and $\langle 0|0\rangle =1.$ The
vectors

\begin{equation}
|\nu _{1},\nu _{-1}\rangle =\frac{(b_{1}^{\dagger })^{\nu
_{1}}(b_{-1}^{\dagger })^{\nu _{-1}}}{\sqrt{\nu _{1}!\nu _{-1}!}}|0\rangle,
\label{cs}
\end{equation}
where $\nu _{1},\nu _{-1}$ run over all non-negative integers, form an
orthonormal basis in $\mathcal{H}$. They are the common eigenvectors of the
boson number operators $N_{1}=b_{1}^{\dagger }b_{1}$, $N_{-1}=b_{-1}^{
\dagger }b_{-1}$ \ and $N=N_{1}+N_{-1}:$ 
\begin{equation}
\begin{tabular}{ll}
$N_{1}|\nu _{1},\nu _{-1}\rangle = \nu _{1}|\nu _{1},\nu _{-1}\rangle ,$ & 
\\ 
$N_{-1}|\nu _{1},\nu _{-1}\rangle = \nu _{-1}|\nu _{1},\nu _{-1}\rangle ,$ & 
\\ 
$N|\nu _{1},\nu _{-1}\rangle =\nu |\nu _{1},\nu _{-1}\rangle,$ & 
\end{tabular}
\label{na}
\end{equation}
where $\nu =\nu _{1}+\nu _{-1}$ and 
\begin{equation}
\begin{array}{lll}
{N}_{i}=N_{i}^{*}, & {[}N_{i},b_{i}^{\dagger }{]}=b_{i}^{\dagger }, & {[}
N_{i},b_{i}{]}=-b_{i}; \ {i=\pm 1.}
\end{array}
\label{cnc}
\end{equation}

The boson representation of \ $sp(4,R)$ is given in a standard way by means
of the operators $F_{i,,j}=b_{i}^{\dagger }b_{j}^{\dagger
},G_{i,,j}=F_{i,,j}^{*}=b_{i}b_{j}$ and $A_{i,,j}=A_{j,i}^{*}=b_{i}^{
\dagger }b_{j}+\frac{1}{2}\delta _{i,j}$ where $i,j=\pm 1$ \cite{bar}. It is
reducible and decomposes into two irreducible representations, each acting
in the subspaces $\mathcal{H}_{+}$ and $\mathcal{H}_{-}$ of $\mathcal{\ H} $
labeled by the eigenvalue of the $sp(4,R)$ invariant operator $P=(-1)^{N}:$

\[
\mathcal{H=H}_{+}\oplus \mathcal{H}_{-},\;P|\varphi _{\pm }\rangle =\pm
|\varphi _{\pm }\rangle ,\;|\varphi _{\pm }\rangle \in \mathcal{H}_{\pm }.
\]
In other words, $\mathcal{H}\ _{+}$ is spanned by the vectors (\ref{cs})
with $\nu =\nu _{1}+\nu _{-1}$ even and $\mathcal{H}_{-}$ \ with $\nu $-odd,
respectively.

The maximal compact subgroup $U(2)$ of $Sp(4,R)$ can be generated by the
Weyl generators $A_{i,j}$ as well as by the well-known equivalent system 
\begin{equation}
\begin{tabular}{llll}
$I_{+}=b_{1}^{\dagger }b_{-1},$ & $I_{-}=I_{+}^{*}=b_{-1}^{\dagger }b_{1},$
& $I_{0}=I_{0}^{*}=\frac{1}{2}(b_{1}^{\dagger }b_{1}-b_{-1}^{\dagger
}b_{-1}),$ & $N$
\end{tabular}
\label{cu2g}
\end{equation}
that satisfies the commutation relations 
\begin{equation}
\begin{tabular}{llll}
${\lbrack }I_{0},I_{\pm }{]}=\pm I_{\pm },$ & ${[}I_{+},I_{-}{]}=2I_{0},$ & $
\left[ N,I_{\pm }\right] =0,$ & $\left[ N,I_{0}\right] =0.$
\end{tabular}
\label{crc}
\end{equation}
The operators $I_{0},I_{\pm }$ close the algebra $su(2)\sim so(3).$\ The
operator $N$ generates $u(1)$ and plays the role of the first order
invariant (\ref{crc}) of $U(2)=SU(2)\otimes U(1)$. Each of the $\mathcal{H}\
_{+}$ and $\mathcal{H}\ _{-}$ subspaces decompose into a direct sum of
eigensubspaces of $N,$ defined by the condition that $\nu $ is fixed: 
\begin{equation}
\begin{array}{cc}
\mathcal{H}\ _{+}=\ \mathrel{\mathop{\oplus }\limits_{\nu }}{\cal H}_{\nu
}^{+}\ , & \mathcal{H}\ _{-}=\ \mathrel{\mathop{\oplus }\limits_{\nu }}{\cal
H} _{\nu}^{-}\ .
\end{array}
\label{ren}
\end{equation}
An irreducible unitary representation (IUR) of $U(2)$ is realized in each $
\mathcal{H}\ _{\nu }^{\pm }$ space.

Another option for labeling the basis vectors (\ref{cs}) is the eigenvalues
of the second order Casimir operator of $SU(2)$, 
\begin{equation}
\mathbf{I}^{2}=\frac{1}{2}(I_{+}I_{-}+I_{-}I_{+})+I_{0}I_{0}=\frac{N}{2}(
\frac{N}{2}+1)\ ,  \label{i2}
\end{equation}
and its third projection, $I_{0}$: 
\[
\begin{tabular}{ll}
$\mathbf{I}^{2}|i,i_{0}\rangle =i(i+1)|i,i_{0}\rangle ,$ & $
I_{0}|i,i_{0}\rangle =i_{0}|i,i_{0}\rangle \ ,$
\end{tabular}
\]
where from (\ref{i2}) $i=\frac{\nu }{2}=\frac{1}{2}(\nu _{1}+\nu _{-1}),$ $
i_{0}$ $=\frac{1}{2}(\nu _{1}-\nu _{-1})$ and 
\begin{equation}
|i,i_{0}\rangle =\frac{(b_{1}^{\dagger })^{(i+i_{0})}(b_{-1}^{\dagger
})^{(i-i_{0})}}{\sqrt{(i+i_{0})!(i-i_{0})!}}|0\rangle \equiv |\nu _{1},\nu
_{-1}\rangle .  \label{js}
\end{equation}
Applying the raising and lowering operators $I_{\pm }$, 
\begin{equation}
I_{\pm }|i,i_{0}\rangle =\sqrt{(i\mp i_{0})(i\pm i_{0}+1)}|i,i_{0}\rangle \ ,
\label{acio}
\end{equation}
to the lowest $|i,-i\rangle $ (highest $|i,i\rangle $) weight state $\nu $
times we obtain all the basis states of a given representation.

We are also interested in the noncompact content of the boson representation
of $sp(4,R).$

\begin{enumerate}
\item  A reducible unitary representation (``ladder series''\cite{tod}) $
u^{0}(1,1)$ \ of the algebra $u(1,1)$ \ with \ Weyl generators $
b_{1}^{\dagger }b_{1},b_{1}^{\dagger }b_{-1}^{\dagger
},-b_{-1}b_{1},-b_{-1}b_{-1}^{\dagger }$ acts in $\mathcal{H}$. The first
order Casimir operator of $U^{0}(1,1)$ \ is essentially the operator $I_{0}$
, 
\[
C_{1}^{0}=b_{1}^{\dagger }b_{1}-b_{-1}b_{-1}^{\dagger }=2I_{0}-1,
\]
and hence the reduction of the ladder series into IUR's (ladders) can be
carried out by using $I_{0}.$ The spaces $\mathcal{H}\ _{\pm }$ decompose
into direct sums of eigenspaces of $I_{0}$ labeled by $i_{0}=\frac{1}{2}(\nu
_{1}-\nu _{-1})$: 
\begin{equation}
\begin{array}{cc}
\mathcal{H}\ _{+}=\stackrel{}{\mathrel{\mathop{\oplus }\limits_{i_{0}}}}
\mathcal{H}_{i_{0}}^{+}\ , & \mathcal{H}\ _{-}=\mathrel{\mathop{\stackrel{}{
\oplus }}\limits_{i_{0}}}{\cal H}_{i_{0}}^{-}\ .
\end{array}
\label{rej0}
\end{equation}
An irreducible representation (a ladder) of $\ $the $u^{0}(1,1)$ is induced
in each $\mathcal{H}_{i_{0}}^{\pm }$ space. The operators $N=N_{1}+N_{-1}$
and $I_{0}=\frac{1}{2}(N_{1}-N_{-1})$ \ can be considered as another
complete set of operators, both diagonal in the basis (\ref{cs}) and
therefore uniquely specifying the states. We can represent this fact with
the pyramids given on Fig.1, where the rows representing the IUR of $U(2)$
are labeled by $\nu $ and the columns representing the ladders of $
u^{0}(1,1) $ by $i_{0}$. Each cell corresponds to one of the states $|\nu
_{1},\nu _{-1}\rangle $ defined by (\ref{cs}). For the $\mathcal{H}\ _{+}$
space we have \vskip 0.2cm \centerline{\bf Figure 1} 
\[
\begin{tabular}{|l|lll|l|lll|}
\hline
$\nu $/i$_{0}$ & 3 & \multicolumn{1}{|l}{2} & \multicolumn{1}{|l|}{1} & 0 & 
-1 & \multicolumn{1}{|l}{-2} & \multicolumn{1}{|l|}{-3} \\ \hline
0 &  &  &  & $|0\rangle $ &  &  &  \\ \cline{4-4}\cline{4-6}
2 &  &  & \multicolumn{1}{|l|}{$|2,0\rangle $} & $|1,1\rangle $ & $
|0,2\rangle $ & \multicolumn{1}{|l}{} &  \\ \cline{3-7}
4 &  & \multicolumn{1}{|l}{$|4,0\rangle $} & \multicolumn{1}{|l|}{$
|3,1\rangle $} & $|2,2\rangle $ & $|1,3\rangle $ & \multicolumn{1}{|l}{$
|0,4\rangle $} & \multicolumn{1}{|l|}{} \\ \cline{2-8}
6 & $\vdots $ & \multicolumn{1}{|l}{$\vdots $} & \multicolumn{1}{|l|}{$
\vdots $} & $\vdots $ & $\vdots $ & \multicolumn{1}{|l}{$\vdots $} & 
\multicolumn{1}{|l|}{$\vdots $}
\end{tabular}
\]
\vskip 0.2cm and for $\mathcal{H}\ _{-}$ \vskip 0.2cm 
\centerline{\bf Figure 2} 
\[
\begin{tabular}{|l|l|l|l|l|l|l|l|l|}
\hline
$\nu $/$i_{0}$ & $\frac{7}{2}$ & $\frac{5}{2}$ & $\frac{3}{2}$ & $\frac{1}{2}
$ & {\tiny -}$\frac{1}{2}$ & {\tiny -}$\frac{3}{2}$ & {\tiny -}$\frac{5}{2}$
& {\tiny -}$\frac{7}{2}$ \\ \hline
$1$ &  &  &  & $|1,0\rangle $ & $|0,1\rangle $ &  &  &  \\ \hline
$3$ &  &  & $|3,0\rangle $ & $|2,1\rangle $ & $|1,2\rangle $ & $|0,3\rangle $
&  &  \\ \hline
$5$ &  & $|5,0\rangle $ & $|4,1\rangle $ & $|3,2\rangle $ & $|2,3\rangle $ & 
$|1,4\rangle $ & $|0,5\rangle $ &  \\ \hline
$\vdots $ & $\vdots $ & $\vdots $ & $\vdots $ & $\vdots $ & $\vdots $ & $
\vdots $ & $\vdots $ & $\vdots $
\end{tabular}
\]
The set of operators $F_{0}\equiv F_{1,-1}=b_{1}^{\dagger }b_{-1}^{\dagger
},G_{0}\equiv G_{1,-1}=b_{1}b_{-1}$ and $A_{0}=\frac{1}{2}(N+1)$ give a
representation $su^{0}(1,1)$ of the $\ su(1,1)$ algebra. They commute in the
following way: 
\[
\begin{tabular}{lll}
$\left[ A_{0},F_{0}\right] =F_{0},$ & $\left[ A_{0},G_{0}\right] =-G_{0},,$
& $\left[ F_{0},G_{0}\right] =-2A_{0}.$
\end{tabular}
\]
By adding the operator $I_{0}$ we obtain the $u^{0}(1,1)=su^{0}(1,1)\oplus
u^{0}(1)$ extension (\ref{cu2g}).

The second order Casimir invariant of this subgroup is given by 
\begin{equation}
C_{2}(SU^{0}(1,1))=(A_{0})^{2}-\frac{1}{2}(F_{0}G_{0}+G_{0}F_{0})=(I_{0}+
\frac{1}{2})(I_{0}-\frac{1}{2}).
\end{equation}
The quadratic equation $(i_{0}+\frac{1}{2})(i_{0}-\frac{1}{2})=\phi (\phi
+1) $ has two solutions for $\phi $: $\phi _{1}=i_{0}-\frac{1}{2}$ and $\phi
_{2}=-i_{0}-\frac{1}{2}.$ Thus the discrete positive series $D^{+}$ of IURs
of $\ su(1,1)$ is realized for the real negative values of $\phi _{1,2}$ 
\cite{bar}. The corresponding spectra of \ $\phi _{1}(i_{0}\leq 0)$ and $
\phi _{2}(i_{0}\geq 0)$ are: 
\begin{eqnarray*}
\begin{tabular}{|llllll|}
\hline
$\phi _{1}$ & $=$ & $-\frac{1}{2},$ & $-1,$ & $-\frac{3}{2},$ & .... \\ 
\hline
$i_{0}$ & $=$ & \multicolumn{1}{c}{$0,$} & \multicolumn{1}{c}{$-\frac{1}{2},$
} & \multicolumn{1}{c}{$-1,$} & .... \\ \hline
\end{tabular}
\\
\begin{tabular}{|llllll|}
\hline
$\phi _{2}$ & $=$ & $-\frac{1}{2},$ & $-1,$ & $-\frac{3}{2},$ & .... \\ 
\hline
$i_{0}$ & $=$ & \multicolumn{1}{c}{$0,$} & \multicolumn{1}{c}{$\frac{1}{2},$}
& \multicolumn{1}{c}{$1,$} & .... \\ \hline
\end{tabular}
\end{eqnarray*}
In the framework of $su^{0}(1,1)$ a degeneracy takes place - the same IUR of 
$su(1,1)$ is realized in $\mathcal{H}_{i_{0}}^{\pm }$ and in $\mathcal{H}
_{-i_{0}}^{\pm }.$ This degeneracy is removed by the operator $I_{0}$ ,
i.e., after the extension of $su^{0}(1,1)$ to $u^{0}(1,1).$ \ In each
representation of $D^{+}$ the spectrum of $A_{0}$ $=\frac{1}{2}(N+1)$ is
given by $\alpha _{0}=\frac{1}{2}(\nu +1)=-\phi _{i},-\phi _{i}+1,-\phi
_{i}+2.....$ ,$i=1,2.$

\item  Next we consider two mutually complementary representations $
su^{+}(1,1)$ and $su^{-}(1,1)$ of the algebra $su(1,1)\subset sp(4,R)$
acting in $\mathcal{H}$. They are given by the operators $F_{\pm }=\frac{1}{2
}F_{\pm 1,\pm 1}$, $G_{\pm }=\frac{1}{2}G_{\pm 1,\pm 1}$ and $A_{\pm }=\frac{
1}{2}(N_{\pm 1}+\frac{1}{2})$, respectively, with commutation relations 
\[
\begin{tabular}{lll}
$\left[ A_{\pm },F_{\pm }\right] =F_{\pm },$ & $\left[ A_{\pm },G_{\pm
}\right] =-G_{\pm },$ & $\left[ F_{\pm },G_{\pm }\right] =-2A_{\pm }.$
\end{tabular}
\]
It is simple to see that each of the generators of $SU^{+}(1,1)$ commutes
with all the generators of the other $SU^{-}(1,1)$ subgroup. The second
order Casimir operators of the $SU^{\pm }(1,1)$ are

\begin{equation}
C_{2}(SU^{\pm }(1,1))=(A_{0}^{\pm })^{2}-\frac{1}{2}(F_{\pm }G_{\pm }+G_{\pm
}F_{\pm })=-\frac{3}{16}.  \label{clkmsu}
\end{equation}

The equation $\phi (\phi +1)=-\frac{3}{16}$ has two solutions: $\phi
_{1}^{\pm }=-\frac{1}{4}$ and $\phi _{2}^{\pm }=-\frac{3}{4}$. Therefore,
two unitary representations from the $D^{+}$series are realized. The
corresponding spectra of the eigenvalues $\alpha _{\pm }=\frac{1}{2}(\nu
_{\pm 1}+\frac{1}{2})$ of the operators $A_{\pm }$ for $\phi _{1}^{\pm }=-
\frac{1}{4}$ are given by 
\begin{equation}
\begin{tabular}{|llllll|}
\hline
$\alpha _{\pm }$ & $=$ & $\frac{1}{4},$ & $\frac{5}{4},$ & $\frac{9}{4},$ & 
.... \\ \hline
$\nu _{\pm 1}$ & $=$ & $0,$ & $2,$ & $4,$ & .... \\ \hline
\end{tabular}
\label{sp01}
\end{equation}

and for $\phi _{2}^{\pm }=-\frac{3}{4}$ by 
\begin{equation}
\begin{tabular}{|llllll|}
\hline
$\alpha _{\pm }$ & $=$ & $\frac{3}{4},$ & $\frac{7}{4},$ & $\frac{11}{4},$ & 
.... \\ \hline
$\nu _{\pm 1}$ & $=$ & $1,$ & $3,$ & $5,$ & .... \\ \hline
\end{tabular}
\label{sp02}
\end{equation}
In this case the addition of the operators $N_{\mp 1}$ considered as
generators of the representations $U^{\mp }(1)$ of the group $U(1)$, extend $
su^{\pm }(1,1)$ to the $u^{\pm }(1,1)=$ $su^{\pm }(1,1)$ $\oplus u^{\mp
}(1). $\ $N_{\mp 1}$ act as the first order Casimir operators of $U^{\pm
}(1,1)$. The spaces $\mathcal{H}\ _{+}$ and $\mathcal{H}\ _{-}$ are
decomposed into direct sums of eigensubspaces of $N_{-1}$and $N_{1}$: 
\begin{equation}
\begin{tabular}{lll}
$\mathcal{H}_{+}=$ & $(\mathrel{\mathop{\stackrel{\infty }{\oplus
}}\limits_{k=0}}{\cal H}_{\nu _{\mp 1}=2k}(\phi =-\frac{1}{4}))\oplus $ & $( 
\mathrel{\mathop{\stackrel{\infty }{\oplus }}\limits_{k=1}}{\cal H}_{\nu
_{\mp 1}=2k+1}(\phi =-\frac{3}{4})),$ \\ 
$\mathcal{H}\ _{-}=$ & $(\mathrel{\mathop{\stackrel{\infty }{\oplus
}}\limits_{k=0}}{\cal H}_{\nu _{\mp 1}=2k}(\phi =-\frac{3}{4}))\oplus $ & $(
\mathrel{\mathop{\stackrel{\infty }{\oplus }}\limits_{k=1}}{\cal H}_{\nu
_{\mp 1}=2k+1}(\phi =-\frac{1}{4})).$
\end{tabular}
\label{dehpm}
\end{equation}
In each $\mathcal{H}\ _{\nu _{\mp 1}}^{\pm }(\phi _{i}),i=1,2$ a IUR\ of $
u(1,1) $ \ is realized. The degeneracy which takes place on the level of $
su(1,1)$ is completely removed after the extension to $u^{\pm }(1,1).$ The
subspaces $\mathcal{H}\ _{\nu _{\mp 1}}^{\pm }(\phi _{i}),i=1,2$ are
represented \ on fig.1 by the diagonals defined by the conditions $\nu _{\mp
1}$ fixed.

\item  Finally we can construct another representation of $su(1,1)$by the
simple sum of the generators of the $SU^{\pm }(1,1)$:

\[
\begin{array}{c}
F=\frac{1}{2}(F_{1,1}+F_{-1,-1}), \\ 
G=\frac{1}{2}(G_{1,1}+G_{-1,-1}), \\ 
A=\frac{1}{2}(N_{1}+N_{-1}+1)\equiv A_{0}.
\end{array}
\]
\end{enumerate}

\section{$q$-bosons and the quantum $sp_{q}(4,R)$ algebra}

In this section, using the $q$-deformation of the classical bosons, we are
going to construct a $q$-deformation $sp_{q}(4,R)$ of the boson
representation of the$\ sp(4,R)$ algebra \cite{Hay}, in the same manner as
in the previous section. We start by deforming the operators $b_{i}^{\dagger
}$ and $b_{i}$, $i=\pm 1$, by means of the transformation \cite{kud}:

\begin{equation}
\begin{array}{ll}
a_{i}^{\dagger }=\sqrt{\frac{[N_{i}]}{N_{i}}}b_{i}^{\dagger }, & a_{i}=\sqrt{
\frac{[N_{i}+1]}{N_{i}+1}}b_{i},
\end{array}
\label{rac}
\end{equation}
where $[X]\equiv \frac{q^{X}-q^{-X}}{q-q^{-1}}$. Obviously $(a_{i}^{\dagger
})^{*}=a_{i}.$ It is possible to interpret the deformation of the classical
boson creation and annihilation operators $b_{i}^{\dagger }$ and $b_{i}$
where $i=\pm 1$, by analyzing the expansion of the coefficients in (\ref{rac}
) in terms of the deformation parameter $\tau ,$ introduced as $q=e^{\tau }$:

\begin{equation}
\frac{\lbrack N_{i}]}{N_{i}}=\allowbreak 1+\frac{1}{6}\left(
N_{i}^{2}-1\right) \tau ^{2}+\frac{1}{12}\left( \frac{1}{10}N_{i}^{4}-\frac{1
}{3}N_{i}^{2}+\frac{7}{30}\right) \tau ^{4}+O\left( \tau ^{6}\right) .
\label{exn}
\end{equation}
In this case we have an infinite expansion containing all the even powers of
the deformation parameter and also all the even powers of each of the
classical operators $N_{i}$ of the number of bosons.

From (\ref{rac}) it is easy to obtain the $q$-deformed commutation relations
for the deformed oscillators:

\begin{equation}
a_{i}a_{i}^{\dag }-q^{1}a_{i}^{\dag }a_{i}=q^{-N_{i}},  \label{ac1}
\end{equation}
\begin{equation}
a_{i}a_{i}^{\dag }-q^{-1}a_{i}^{\dag }a_{i}=q^{N_{i}},  \label{ac2}
\end{equation}
\[
\begin{array}{ll}
\lbrack a_{i},a_{k}^{\dagger }]=0, & [a_{i}^{\dagger },a_{k}^{\dagger
}]=[a_{i},a_{k}]=0,i\neq k.
\end{array}
\]

In terms of the deformed boson operators, the basis vectors (\ref{cs}) are ($
a_{i}|0\rangle =0)$: 
\begin{equation}
|\nu _{1},\nu _{-1}\rangle =\frac{(a_{1}^{\dagger })^{\nu
_{1}}(a_{-1}^{\dagger })^{\nu _{-1}}}{\sqrt{[\nu _{1}]![\nu _{-1}]!}}
|0\rangle \equiv \frac{(b_{1}^{\dagger })^{\nu _{1}}(b_{-1}^{\dagger })^{\nu
_{-1}}}{\sqrt{\nu _{1}!\nu _{-1}!}}|0\rangle ,  \label{qs}
\end{equation}
where $[X]!=[1][2][3]....[X].$ Obviously the spectrum of the operators $
N_{i} $, $i=\pm 1$, is preserved (\ref{na}). It is easy to see that their
relations with the $q$-deformed bosons are the same as (\ref{cnc}): 
\begin{equation}
\begin{array}{ll}
{\lbrack }N_{i},a_{i}{]}=-a_{i}{\ ,} & {[}N_{i},a_{i}^{\dag }{]}=a_{i}^{\dag
}.
\end{array}
\label{c2}
\end{equation}

A $q$-boson representation of a $q$-deformed algebra $sp_{q}(4,R)$ acts in
the Fock space $\mathcal{H}$, which can be represented by the following set
of operators: 
\begin{equation}
\begin{tabular}{lll}
$F_{i,j}^{q}=a_{i}^{\dagger }a_{j}^{\dagger },$ & $
G_{i,j}^{q}=(F_{j,i}^{q})^{*}=a_{i}a_{j},$ & $i,j=\pm 1,$ \\ 
$J_{\pm }=a_{\pm 1}^{\dagger }a_{\mp 1},$ & $J_{0}=\frac{1}{2}
(N_{1}-N_{-1})\equiv I_{0},$ & $N=N_{1}+N_{-1}.$
\end{tabular}
\label{gqa}
\end{equation}
In this representation the raising and lowering operators $
\;F_{i,j}^{q},\;G_{i,j}^{q}$ and $J_{+},\;J_{-}=J_{+}^{*}$ are deformed. The
complete set of operators $N$ and $I_{0}$ used in the $sp(4,R)$ case is
retained after the deformation. This allows one to rewrite the basic states (
\ref{qs}) in the form (compare with (\ref{js})): 
\[
|\nu _{1},\nu _{-1}\rangle ==\frac{(a_{1}^{\dagger })^{j+m}(a_{-1}^{\dagger
})^{j-m}}{\sqrt{[j+m]![j-m]!}}|0\rangle ,
\]
where 
\[
\begin{tabular}{ll}
$N|j,m\rangle =2j|j,m\rangle,$ & $J_{0}|j,m\rangle =m|j,m\rangle; \ \
(j\equiv i,m\equiv i_{0}).$
\end{tabular}
\]
It can be checked directly that 
\[
\begin{tabular}{ll}
$\left[ J_{\pm },N\right] =0,$ & $\left[ J_{0},N\right] =0,$ \\ 
$\left[ F_{i,j}^{q},N\right] =-2F_{i,j}^{q},$ & $\left[ G_{i,j}^{q},N\right]
=2G_{i,j}^{q}.$
\end{tabular}
\]
It follows that in this case the operator $P=(-1)^{N}$ also commutes with
all the elements of the representation of the $sp_{q}(4,R)$ considered. In
other words, the decomposition \ $\mathcal{H=H}_{+}\oplus \mathcal{H}_{-}$ \
remains the same after the deformation. Thus the $q$-boson $sp_{q}(4,R)$
representation decomposes into two irreducible ones acting in $\mathcal{H}
_{+}$ and $\mathcal{H}_{-}$, respectively.

The reduction to subalgebras is the same as in the nondeformed case. In $
\mathcal{H}$ there is a reducible representation of $u_{q}(2)$ given by $N$
and the operators $J_{\pm },J_{0}$ which commute in the following way: 
\begin{equation}
\begin{tabular}{ll}
$\left[ J_{0},J_{\pm }\right] =\pm J_{\pm },$ & $\left[ J_{+},J_{-}\right]
=[2J_{0}].$
\end{tabular}
\label{cruq2}
\end{equation}
Since the same operator $N$ acts also as a first order invariant of $
u_{q}(2) $, the decomposition (\ref{ren}) of the spaces $\mathcal{H}_{+}$
and $\mathcal{H} _{-}$ remains after the deformation. In each of the $
\mathcal{H}_{\nu }^{\pm }$ spaces there acts as well an IUR of $su_{q}(2)$
generated by $J_{0},J_{\pm }.$ The second order Casimir operator in this
case is given by the operator 
\begin{eqnarray}
\mathbf{J}^{2} &=&\frac{1}{2}(J_{+}J_{-}+J_{-}J_{+})+\frac{1}{2}
([J_{0}][J_{0}+1]+[J_{0}][J_{0}-1])=  \label{c2uq2} \\
&=&J_{+}J_{-}+[J_{0}][J_{0}-1]=J_{-}J_{+}+[J_{0}][J_{0}+1]=\left[ \frac{N}{2}
\right] \left[ \frac{N}{2}+1\right] .  \nonumber
\end{eqnarray}
Its eigenvalues equation has the form 
\[
\mathbf{J}^{2}|j,m\rangle =[j][j+1]|j,m\rangle .
\]
The action of the deformed $J_{\pm }$ is: 
\begin{equation}
J_{\pm }|j,m\rangle =\sqrt{[j\pm m+1][j\mp m]}|j,m\rangle .  \label{ajpm}
\end{equation}
By acting with $J_{\mp }$ on the highest ( lowest) weight states $
|j,j\rangle \;(|j,-j\rangle )$ $2j$ times we obtain all the basis in the
space $\mathcal{\ H}_{\nu }^{\pm }$ of a given IUR of $su_{q}(2)$ (a row in
the diagrams in Fig.1). So in the context of the boson $sp_{q}(4,R)$
representation, we have a full description of IURs of the deformed $
su_{q}(2) $ algebra.

Thus far we have focused on compact structures; we now turn to a
consideration of noncompact cases.

\begin{enumerate}
\item  In the $\mathcal{H}$ space, a deformation $u_{q}^{0}(1,1)$ \cite{kud}
of the $u^{0}(1,1)$ algebra acts. It is generated by the operators 
\[
\begin{tabular}{ll}
$K_{+}^{0}=F_{1,-1}^{q}=a_{1}^{\dagger }a_{-1}^{\dagger }$, & $
K_{-}^{0}=G_{1,-1}^{q}=a_{1}a_{-1},$ \\ 
$K_{0}^{0}=\frac{1}{2}(N+1),$ & $J_{0},$
\end{tabular}
\]
which have the following commutation rules: 
\[
\begin{tabular}{ll}
$\left[ K_{0}^{0},K_{\pm }^{0}\right] =\pm K_{\pm }^{0},$ & $\left[
K_{+}^{0},K_{-}^{0}\right] =-[2K_{0}^{0}],$ \\ 
$\left[ K_{0}^{0},J_{0}\right] =0,$ & $\left[ K_{\pm }^{0},J_{0}\right] =0.$
\end{tabular}
\]
The reduction of $\mathcal{H}$ to eigenspaces of the operator $J_{0}=I_{0}$ (
\ref{rej0}) is invariant with respect to the deformation. So the IURs
(ladders) of \ $u_{q}^{0}(1,1)$ act in $\mathcal{H}_{m}^{\pm }$, $m\equiv
i_{0}$. IURs of $su_{q}^{0}(1,1)\subset u_{q}^{0}(1,1)$ and generated by $
K_{\pm }^{0},K_{0}^{0}$ act also in $\mathcal{H}_{m}^{\pm }$. The second
order Casimir invariant of $SU_{q}^{0}(1,1)$ is given by: 
\begin{equation}
\begin{tabular}{lll}
$(\mathbf{K}^{0})^{2}=$ & $\frac{1}{2}
([K_{0}^{0}][K_{0}^{0}+1]+[K_{0}^{0}][K_{0}^{0}-1])$ & $-\frac{1}{2}
(K_{+}^{0}K_{-}^{0}+K_{-}^{0}K_{+}^{0})\equiv $ \\ 
& $\equiv [K_{0}^{0}][K_{0}^{0}+1]-K_{-}^{0}K_{+}^{0}\equiv $ &  \\ 
& $\equiv [K_{0}^{0}][K_{0}^{0}-1]-K_{+}^{0}K_{-}^{0}=$ & $\left[
J_{0}\right] ^{2}-\left[ \frac{1}{2}\right] ^{2}.$
\end{tabular}
\label{qc2u11}
\end{equation}
The last expression is obtained with the help of the relations: 
\begin{eqnarray*}
K_{+}^{0}K_{-}^{0} &=&\left[ \frac{N}{2}\right] ^{2}-\left[ J_{0}\right]
^{2}, \\
\left[ K_{0}^{0}\right] \left[ K_{0}^{0}-1\right] &=&\left[ \frac{N}{2}
\right] ^{2}-\left[ \frac{1}{2}\right] ^{2}.
\end{eqnarray*}
Hence the eigenvalues of $(\mathbf{K}^{0})^{2}$ on the basis vectors are 
\[
(\mathbf{K}^{0})^{2}|j,m\rangle =([m]+\left[ \frac{1}{2}\right] )([m]-\left[ 
\frac{1}{2}\right] )|j,m\rangle .
\]
The equation 
\[
([m]+\left[ \frac{1}{2}\right] )([m]-\left[ \frac{1}{2}\right] )=\phi
^{q}(\phi ^{q}+2\left[ \frac{1}{2}\right] )
\]
has two solutions for $\phi ^{q}$: $\phi _{1}^{q}=[m]-\left[ \frac{1}{2}
\right] $ and $\phi _{2}^{q}=-[m]-\left[ \frac{1}{2}\right] .$ Imposing $
\phi ^{q}<0$ we obtain for $q>0$ a description of the deformed discrete
series $D_{q}^{+}$ of $su_{q}^{0}(1,1).$ The spectra simultaneously run by $
\phi _{i}^{q}$ and $m$ are, respectively, for $\phi _{1}^{q}<0$ $(m\leq 0)$: 
\[
\begin{tabular}{|l|l|l|l|l|}
\hline
$\phi _{1}^{q}=$ & $-\left[ \frac{1}{2}\right] $ & $-2\left[ \frac{1}{2}
\right] $ & $-[1]-\left[ \frac{1}{2}\right] $ & $-\left[ \frac{3}{2}\right]
-\left[ \frac{1}{2}\right] .....$ \\ \hline
$m=$ & \multicolumn{1}{|c|}{$0$} & \multicolumn{1}{|c|}{$-\frac{1}{2}$} & 
\multicolumn{1}{|c|}{$-1$} & \multicolumn{1}{|c|}{$-\frac{3}{2}.......$} \\ 
\hline
\end{tabular}
\]
and for $\phi _{2}^{q}<0$ $(m\geq 0)$: 
\[
\begin{tabular}{|l|l|l|l|l|}
\hline
$\phi _{1}^{q}=$ & $-\left[ \frac{1}{2}\right] $ & $-2\left[ \frac{1}{2}
\right] $ & $-[1]-\left[ \frac{1}{2}\right] $ & $-\left[ \frac{3}{2}\right]
-\left[ \frac{1}{2}\right] .....$ \\ \hline
$m=$ & \multicolumn{1}{|c|}{$0$} & \multicolumn{1}{|c|}{$\frac{1}{2}$} & 
\multicolumn{1}{|c|}{$1$} & \multicolumn{1}{|c|}{$\frac{3}{2}.......$} \\ 
\hline
\end{tabular}
\]
A degeneracy of the same type as in the nondeformed case takes place because
of the quadratic dependence on $[J_{0}]$ in (\ref{qc2u11}). But we should
underline that in this case also the degeneracy is removed by the operator $
J_{0},$ which remains nondeformed and is acting as a first invariant of $
u_{q}^{0}(1,1)$. The spectrum of $K_{0}^{0}$ $=\frac{1}{2}(N+1)$, which is
also a nondeformed operator, is related to the nondeformed $\phi $ $(\phi
^{q}\mathrel{\mathop{\rightarrow }\limits_{q\rightarrow 1}}\phi )$, so we
have: 
\[
\frac{1}{2}(\nu +1)=-\phi ,-\phi +1,-\phi +2,.......
\]

\item  At the end we will discuss the deformations $u_{q}^{\pm }(1,1)$ \cite
{kud} of the two mutually complementary representations $u^{\pm }(1,1)$,
each realized by only one kind of $q$-boson. The operators 
\begin{equation}
\begin{tabular}{ll}
$K_{+}^{\pm }=\frac{1}{[2]}F_{\pm 1,\pm 1}^{q}=\frac{1}{[2]}a_{\pm
1}^{\dagger }a_{\pm 1}^{\dagger },$ & $K_{-}^{\pm }=\frac{1}{[2]}G_{\pm
1,\pm 1}^{q}=\frac{1}{[2]}a_{\pm 1}a_{\pm 1},$ \\ 
$K_{0}^{\pm }=\frac{1}{2}(N_{\pm 1}+\frac{1}{2}),$ & $N_{\mp 1}$
\end{tabular}
\end{equation}

commute among themselves in the following way: 
\[
\begin{tabular}{ll}
$\left[ K_{0}^{\pm },K_{\pm }^{\pm }\right] =\pm K_{\pm }^{\pm },$ & $\left[
K_{+}^{\pm },K_{-}^{\pm }\right] =-[2K_{0}^{\pm }]_{2},$ \\ 
$\left[ K_{0}^{\pm },N_{\mp 1}\right] =0,$ & $\left[ K_{\pm }^{\pm },N_{\mp
1}\right] =0,$
\end{tabular}
\]
where the notation $[X]_{m}\equiv \frac{q^{mX}-q^{-mX}}{q^{m}-q^{-m}}$
applies. The nondeformed operators $N_{\mp 1}$ extend the $su_{q}^{\pm
}(1,1) $ to $u_{q}^{\pm }(1,1)$ and act as first-order Casimir invariants.
The second-order Casimir invariants have a slightly modified form (compare
with (\ref{qc2u11})): 
\begin{eqnarray*}
C_{2}\{SU_{q}^{\pm }(1,1)\} &=&[K_{0}^{\pm }]_{2}[K_{0}^{\pm
}+1]_{2}-K_{-}^{\pm }K_{+}^{\pm }= \\
&=&[K_{0}^{\pm }]_{2}[K_{0}^{\pm }-1]_{2}-K_{+}^{\pm }K_{-}^{\pm }.
\end{eqnarray*}
In this case we have the following expressions: 
\begin{eqnarray*}
K_{-}^{\pm }K_{+}^{\pm } &=&\frac{1}{[2]^{2}}\left[ N_{\pm }+1\right] \left[
N_{\pm }+2\right] , \\
K_{+}^{\pm }K_{-}^{\pm } &=&\frac{1}{[2]^{2}}\left[ N_{\pm }\right] \left[
N_{\pm }-1\right] \\
&=&\frac{1}{[2]^{2}}\{\left[ \frac{1}{2}(2N_{\pm }-1)\right] ^{2}-\left[ 
\frac{1}{2}\right] ^{2}\}, \\
\lbrack K_{0}^{\pm }]_{2}[K_{0}^{\pm }-1]_{2} &=&\frac{1}{[2]^{2}}\{\left[ 
\frac{1}{2}(2N_{\pm }-1)\right] ^{2}-1\}.
\end{eqnarray*}
As a result we obtain: 
\begin{eqnarray*}
C_{2}\{SU_{q}^{\pm }(1,1)\} &=&(\mathbf{K}^{\pm })^{2}=\frac{1}{[2]^{2}}
\left( \left[ \frac{1}{2}\right] ^{2}-1\right) \\
&=&\left( \left[ \frac{1}{4}\right] _{2}-\left[ \frac{1}{2}\right]
_{2}\right) \left( \left[ \frac{1}{4}\right] _{2}+\left[ \frac{1}{2}\right]
_{2}\right) .
\end{eqnarray*}
In this case the $q$-deformed equation 
\[
(\mathbf{K}^{\pm })^{2}=^{q}\varphi ^{\pm }(^{q}\varphi ^{\pm }+2\left[ 
\frac{1}{2}\right] _{2})
\]
has solutions 
\[
\begin{tabular}{ll}
$^{q}\varphi _{1}^{\pm }=\left[ \frac{1}{4}\right] _{2}-\left[ \frac{1}{2}
\right] _{2},$ & $^{q}\varphi _{2}^{\pm }=-\left[ \frac{1}{4}\right]
_{2}-\left[ \frac{1}{2}\right] _{2},$
\end{tabular}
\]
which means that in the $\mathcal{H}$ space a discrete series of two
$q$-deformed representations of \ $su_{q}(1,1)$ is realized for each kind of
$q$-boson (with index $+1$or $-1)$. The spectra of $K_{0}^{\pm }$ correspond
to the limit $^{q}\varphi _{i}^{\pm }\mathrel{\mathop{\rightarrow }
\limits_{q\rightarrow 1}}\varphi _{i}^{\pm }$, $i=1,2$ and coincide with the
spectra of the operators $A_{\pm }$ in the nondeformed picture (see (\ref
{sp01}) and (\ref{sp02})). Here, again the extension from $su_{q}^{\pm
}(1,1) $ to $u_{q}^{\pm }(1,1)=su_{q}^{\pm }(1,1)\oplus u^{\mp }(1)$ is
realized by adding the operators $N_{\mp 1}$. Thus the degeneracy is
eliminated in the same way as in the nondeformed case. Now the spaces in
which the irreducible boson representations of the classical $u^{\pm }(1,1)$
and $q$-deformed $u_{q}^{\pm }(1,1)$ act are the same and the decompositions
(\ref{dehpm}) are the same, as we have 
\begin{eqnarray*}
\mathcal{H}_{\nu _{\mp }}\left( ^{q}\varphi _{1}^{\pm }=-\left[ \frac{1}{4}
\right] _{2}-\left[ \frac{1}{2}\right] _{2}\right) &\equiv &\mathcal{H}_{\nu
_{\mp }}\left( \alpha _{1}^{\pm }=-\frac{3}{4}\right) , \\
\mathcal{H}_{\nu _{\mp }}\left( ^{q}\varphi _{2}^{\pm }=\left[ \frac{1}{4}
\right] _{2}-\left[ \frac{1}{2}\right] _{2}\right) &\equiv &\mathcal{H}_{\nu
_{\mp }}\left( \alpha _{2}^{\pm }=-\frac{1}{4}\right) .
\end{eqnarray*}
\end{enumerate}

\section{Deformation in terms of $su_{q}(2)$ tensor operators}

The $q$-deformed bosons $a_{i}^{\dagger }$ and $a_{i}$, $i=\pm 1$, are not
components of tensor operators with respect to the standard $su_{q}(2)$
defined in the previous section \cite{bid,6}. However, the following
nontrivial modification of the creation and annihilation operators~for $
k=\pm 1$ introduced in eqs. (\ref{rac}) and (\ref{ac2}),

\begin{eqnarray}
t_{k}^{\dagger } &=&q^{\frac{k}{4}}a_{k}^{\dag }q^{\frac{kN_{-k}}{2}}\ \ ;
\label{sp1} \\
t_{k} &=&q^{-\frac{k}{4}}a_{-k}q^{\frac{-kN_{k}}{2}}  \label{sp2}
\end{eqnarray}
transforms these operators \cite{7b} into two-dimensional conjugated,
spinor-like (of rank $\frac{1}{2})$ tensors, $\ (t_{k}^{\dagger})^{\ast
}=t_{-k}$, with respect to $su_{q}(2)$. These deformations can be related to
the classical bosons $b_{i}^{\dagger }$, $b_{i}$, $i=\pm 1$, by means of
their transformations to $q$-deformed oscillators (\ref{rac}).

From the oscillator commutation relations (\ref{ac1}) and (\ref{ac2}) we
obtain the commutation relations: 
\begin{equation}
\begin{array}{ll}
\lbrack \ t_{k},\ t_{l}^{\dagger }]_{q^{\rho }}= & q^{-\frac{k}{2}}\delta
_{k,-l}q^{-2J_{0}}, \\ 
\lbrack \ t_{k}^{\dagger },\ t_{l}^{\dagger }]_{q^{\sigma }}= & [\ t_{k},\
t_{l}]_{q^{\sigma }}=0, \\ 
\rho =\frac{l-k}{2}=-\sigma , & l,k=\pm 1.
\end{array}
\label{cb}
\end{equation}
In this case another $q$-deformation, $sp_{t}(4,R)$, of the $sp(4,R)$ algebra,
with generalized Gauss decomposition 
\[
\mathcal{G}=g_{-}\oplus h\oplus g_{+},
\]
is constructed by the tensor products of the fundamental oscillator
representation for $su_{q}(2)$ (\ref{sp1}) and (\ref{sp2}):

\begin{eqnarray}
(\ t\ \otimes t\ )_{m}^{l}:=T_{m}^{l}\ \supseteq g_{+}\ , &&l\ =\ 1,m=0,\pm
1,  \label{a0} \\
(\ t^{\dagger }\ \otimes t^{\dagger }\ )_{m}^{l}:=\tilde{T}_{m}^{l}\
\supseteq g_{-},\ &&l\ =\ 1,m=0,\pm 1,  \label{a1} \\
(\ t^{\dagger }\ \otimes \ \ t\ )_{m}^{l}:=L_{m}^{l}\ \supseteq h,\ &&l\ =\
0,1,m=-l,-l+1,...,l.  \label{a3}
\end{eqnarray}

This $sp_{t}(4,R)$ algebra \cite{7b} is decomposed in a natural way into a
deformed compact subalgebra $h=su_{t}(2)\otimes u_{t}(1)$ that is generated
by the spherical tensors $L_{m}^{1}$ ($m=0,\pm 1$) and $L_{0}^{0}$ (\ref{a3}
) and ${g}_{+}$ and ${g}_{-}$, which are two $q$-nilpotent subalgebras
containing the components of the two conjugated first rank tensors $
T_{m}^{1} $ ($m=0, \pm 1$) (\ref{a0}) and $\tilde{T}_{m}^{1}$ ($m=0, \pm 1$)
(\ref{a1} ). In this case the $su_{t}(2)$ generated by the components of a
first rank tensor $L_{m}^{1}$ ($m=0,\pm 1$) (\ref{a3}) can be interpreted 
\cite{mesa93} as isomorphic by construction to a deformation of $so(3)$ --
the classical algebra of the angular momentum. Using the $q$-deformed
realization \cite{6} of the Clebsh-Gordon coefficients for $su_{q}(2)$ (\ref
{gqa}), we obtain the explicit expressions for the operators (\ref{a0}), (
\ref{a1}) and (\ref{a3}) in terms of the $q$-spinors (\ref{sp1}) and (\ref
{sp2}):

\begin{equation}
\begin{tabular}{lll}
$T_{1}^{1}=$ & $t_{1}^{\dagger }t_{1}^{\dagger }=$ & $(\tilde{T}
_{-1}^{1})^{\ast },$ \\ 
$T_{-1}^{1}=$ & $t_{-1}^{\dagger }t_{-1}^{\dagger }=$ & $(\tilde{T}
_{1}^{1})^{\ast },$ \\ 
$T_{0}^{1}=$ & $q^{-\frac{1}{2}}\sqrt{[2]}t_{1}^{\dagger }t_{-1}^{\dagger }=$
& $(\tilde{T}_{0}^{1})^{\ast },$
\end{tabular}
\label{rTc}
\end{equation}

\begin{equation}
\begin{tabular}{lll}
$L_{1}^{1}=$ & $t_{1}^{\dagger }t_{1}=q^{-\frac{1}{2}}J_{+}q^{-J_{0}}$ & $
=(L_{-1}^{1})^{*},$ \\ 
$L_{-1}^{1}=$ & $t_{-1}^{\dagger }t_{-1}=q^{\frac{1}{2}}J_{-}q^{-J_{0}}$ & $
=(L_{1}^{1})^{*}.$
\end{tabular}
\label{rLc}
\end{equation}
The above eight operators are the $q$-tensor analogues of the deformed
raising and lowering generators of the $sp_{q}(4,R).$Further we consider the
operators:

\begin{equation}
\begin{tabular}{ll}
$\mathcal{N}_{1}=t_{1}^{\dagger }t_{-1}=q^{\frac{1}{2}}[N_{1}]q^{N_{-1}},$ & 
$\mathcal{N}_{-1}=t_{-1}^{\dagger }t_{1}=q^{-\frac{1}{2}}[N_{-1}]q^{-N_{1}},$
\\ 
$t_{-1}t_{1}^{\dagger }=q^{\frac{1}{2}}[N_{1}+1]q^{N_{-1}},$ & $
t_{1}t_{-1}^{\dagger }=q^{-\frac{1}{2}}[N_{-1}+1]q^{-N_{1}},$
\end{tabular}
\label{wcg}
\end{equation}
obtained by means of the substitutions (\ref{sp1}) and (\ref{sp2}). It must
be noted that in this case those are deformed operators and do not have
expression in terms of the classical bosons unlike the boson number
operators $N_{1}$and $N_{-1},$ used in the case of $sp_{q}(4,R).$

By means of an expansion like that introduced in (\ref{exn}) it is easy to
verify the mixing of two kinds of oscillators $k=\pm 1$ introduced through
the use of tensor operators 
\begin{eqnarray}
\left[ N_{k}\right] \ q^{\pm N_{-k}} &=&N_{k}\pm N_{k}N_{-k}\tau +\frac{1}{6}
N_{k}\left( 3N_{-k}^{2}+N_{k}^{2}-1\right) \tau ^{2}  \label{qnm} \\
&&\pm \frac{1}{6}N_{k}N_{-k}\left( N_{k}^{2}-1+N_{-k}^{2}\right) \tau
^{3}+O\left( \tau ^{4}\right) .  \nonumber
\end{eqnarray}
It is simple to see that the operators $\mathcal{N}_{1}$ and $\mathcal{N}
_{-1}$ belong to the enveloping algebra of the classical oscillators. In (
\ref{qnm}) all the powers of the deformation parameter $\tau $ and all the
degrees of the two ``classical'' operators ${N}_{k}=b_{k}^{\dagger }b_{k}$
for $k=\pm 1$ appear. Using (\ref{rac}), (\ref{sp1}), and (\ref{sp2}) we
have the following relations: 
\[
\begin{tabular}{ll}
$\left[ {N}_{k},t_{k}^{\dagger }\right] =t_{k}^{\dagger },$ & $\left[ {N}
_{k},t_{-k}\right] =-t_{-k},$ \\ 
$\left[ {N}_{k},t_{-k}^{\dagger }\right] =0,$ & $\left[ {N}_{k},t_{k}\right]
=0,$
\end{tabular}
\]
and as a result we have the correct tensor properties: 
\[
\begin{tabular}{ll}
$\left[ {J}_{0},t_{k}^{\dagger }\right] =\frac{k}{2}t_{k}^{\dagger },$ & $
\left[ {J}_{0},t_{k}\right] =\frac{k}{2}t_{k}.$
\end{tabular}
\]

The third component $L_{0}^{1}$ and scalar operator $L_{0}^{0}$ are obtained
as:

\begin{equation}
\begin{tabular}{ll}
$L_{0}^{1}=$ & $\frac{1}{[2]}(q[N_{1}]q^{N_{-1}}-q^{-1}[N_{-1}]q^{-N_{1}})$
\\ 
\ \ \ \ $=$ & $\frac{1}{[2]}(q[N_{1}][N_{-1}+1]-q^{-1}[N_{-1}][N_{1}+1]),$
\end{tabular}
\label{l0r}
\end{equation}
\[
L_{0}^{0}=([N_{1}]q^{N_{-1}}+[N_{-1}]q^{-N_{1}})=[N].
\]
Using (\ref{cb}) we find the commutation relations (\ref{a0}), (\ref{a1})
and (\ref{a3}):

\begin{equation}
{\lbrack }L_{+1}^{1},L_{-1}^{1}{]}=[2]L_{0}^{1}q^{-2J_{0}};{[}
L_{0}^{1},L_{\pm 1}^{1}{]}=\pm q^{\mp 1}L_{\pm 1}^{1}q^{-2J_{0}} .
\label{d12}
\end{equation}
From (\ref{d12}) it is obvious that the components of the first-rank tensors 
$L_{m}^{1}$ ($m=0,\pm 1$) close in a natural way on another deformation, $
su_{t}(2)$, of the classical $su(2)$. The scalar operator $L_{0}^{0}=[N]$
commutes with all the components of the first-rank tensors $L_{m}^{1}$ ($
m=0,\pm 1$), 
\[
\left[ \lbrack N],L_{m}^{1}\right] =0,
\]
and yields decomposition $u_{t}(2)$ $=su_{t}(2)\oplus u_{q}(1)$ with
first-order Casimir invariant $[N]$. The second-order Casimir operator for $
SU_{t}(2)$ is calculated as the scalar product: 
\begin{equation}
-\sqrt{[3]}(\mathbf{L}\otimes \mathbf{L}
)^{0}=qL_{-1}L_{+1}+q^{-1}L_{+1}L_{-1}-L_{0}L_{0}=\frac{1}{[2]}[N][N+2].
\label{l2}
\end{equation}

For completeness we present all the other commutation relations of the
tensor operators (\ref{a0}), (\ref{a1}), and (\ref{a3}) in a slightly
different form than in \cite{7a}. The commutators of $L_{\pm 1}^{1}$ with
the pair raising and lowering operators define their transformation
properties in respect to the $q$-deformed $so(3)$ subalgebra:

\begin{equation}
\begin{array}{ll}
{\lbrack }L_{\pm 1}^{1},\tilde{T}_{m}^{1}{]}= & \pm q^{-(m\pm 1)}\sqrt{{[}
1\mp m{]}{[}1\pm m+1{]}}\tilde{T}_{m\pm 1}^{1}q^{-2J_{0}}, \\ 
{\lbrack }L_{\pm 1}^{1},T_{m}^{1}{]}= & \pm q^{-(m\pm 2)}\sqrt{{[}1\mp m{]}{[
}1\pm m+1{]}}T_{m\pm 1}^{1}q^{-2J_{0}}.
\end{array}
\label{ntp}
\end{equation}
Hence the operators $\tilde{T}_{m}^{1}$ and $T_{m}^{1},m=0,\pm 1$ form two
conjugated vectors with respect to the $so(3)$ subalgebra. From the
following commutators for $k=\pm 1$ and $m=\pm 1$:

\begin{equation}
\begin{array}{ll}
\lbrack \tilde{T}_{m}^{1},\mathcal{N}_{k}]_{q^{2k}}=\delta _{m,-k}q^{\frac{3
}{2}k}[2]\tilde{T}_{m}^{1}q^{-2J_{0}}, & [\mathcal{N}
_{k},T_{m}^{1}]_{q^{2k}}=\delta _{m,k}q^{-\frac{k}{2}
}[2]T_{m}^{1}q^{-2J_{0}}, \\ 
{\lbrack }\tilde{T}_{0}^{1},\mathcal{N}_{k}]_{q^{2k}}{\ =}q^{\frac{k}{2}}
\tilde{T}_{0}^{1}q^{-2J_{0}}, & {[}\mathcal{N}_{k},T_{0}^{1}]_{q^{2k}}{\ =}
q^{\frac{k}{2}}T_{0}^{1}q^{-2J_{0}}
\end{array}
\label{nt}
\end{equation}
it is easy to obtain the commutation relations of $\tilde{T}_{m}^{1}$ and $
T_{m}^{1}$ ($m=0,\pm 1)$ with the operators (\ref{l0r}).

The pair operators $T_{m}^{1}$, $\tilde{T}_{m}^{1}$ generate the two $q$
-nilpotent algebras $g_{+}$, $g_{-}$ and fulfill the following $q$
-commutation relations: 
\begin{equation}
\lbrack T_{m_{1}}^{1}~,~T_{m_{2}}^{1}]_{q^{2(m_{1}-m_{2})}}=0\;,\quad \ [
\tilde{T}_{m_{1}}^{1}~,~\tilde{T}_{m_{2}}^{1}]_{q^{2(m_{1}-m_{2})}}=0\;.
\label{c17}
\end{equation}
The commutation relations between the $T_{m}^{1}$ and $\tilde{T}_{m}^{1}$
close in terms of components of the angular momentum $q$-analog (\ref{a1}).
The subset with $m_{1}+m_{2}\neq 0$ can be presented in a unified way as 
\begin{equation}
\lbrack T_{m_{1}}^{1},\tilde{T}
_{m_{2}}^{1}]_{q^{2(m_{2}-m_{1})}}=-q^{-m_{1}} \sqrt{{[}2{]}}[2(m_{2}-m_{1})
]L_{m_{1}+m_{2}}^{1}q^{-2J_{0}}\;,  \label{c19l}
\end{equation}
while for $m_{1}+m_{2}=0$ we obtain 
\begin{equation}
\begin{array}{lcl}
\lbrack T_{1}^{1},\tilde{T}_{-1}^{1}]_{q^{-4}} & = & -{[}2{]\{q}
^{-2}q^{-2J_{0}}+q^{\frac{1}{2}}{[}2{]\mathcal{N}}_{1}\}q^{-2J_{0}}, \\ 
\lbrack T_{0}^{1},\tilde{T}_{0}^{1}] & = & {[}2{][N+1]}q^{-2J_{0}}, \\ 
\lbrack T_{-1}^{1},\tilde{T}_{1}^{1}]_{q^{4}} & = & -{[}2{]\{q}
^{2}q^{-2J_{0}}+q^{-\frac{1}{2}}[2]\mathcal{N}_{-1}\}q^{-2J_{0}}\;.
\end{array}
\label{c19s}
\end{equation}

In the limit $q\rightarrow 1$ these reproduce the commutation relations of
the ``classical'' $sp(4,R)$ algebra, which has a lot of interesting
applications in the nuclear physics. It should be noted that in this case we
do not obtain a simple generalization of the noncompact $su^{\varepsilon
}(1,1)$ ($\varepsilon =0,\pm $) subalgebras of $sp(4,R)$ as is the case with
the $q$-bosons. By analyzing relations (\ref{nt}) and (\ref{c19s}) it can be
seen that they close in the enveloping algebras of the respective classical $
u^{\varepsilon }(1,1)$ ($\varepsilon =0,\pm $). Actually this is a general
property of this tensor deformation of $sp(4,R)$ because of the appearance
of the $q^{-2J_{0}}\mathrel{\mathop{\rightarrow }\limits _{q\rightarrow 1}}1$
\ factors on the right-hand-side of all commutation relations. The problem
of eliminating this is solved in \cite{7b}.

Working in terms of tensor operators makes the evaluation of the most
general $sp_{t}(4,R)$ invariant operator with respect to the $q$-deformed $
so(3)$ subalgebra quite simple. It is constructed as a linear combination of
the scalar products of (\ref{a0}), (\ref{a1}) and (\ref{a3}) that preserve $
J_{0}$: 
\begin{eqnarray}
S_{2} &=&s_{1}T_{k}^{1}\ .\tilde{T}_{-k}^{1}+s_{2}\tilde{T}
_{k}^{1}.T_{-k}^{1}+s_{3}L_{k}^{1}.L_{-k}^{1}+s_{4}\left[ N\right] ^{2}=
\label{binv} \\
&&s_{1}\left[ N\right] \left[ N-1\right] +s_{2}\left[ N+2\right] \left[
N+3\right] +s_{3}\left[ N\right] \left[ N+2\right] +s_{4}\left[ N\right]
^{2}.  \nonumber
\end{eqnarray}
From this expression it is clear that four additional phenomenological
parameters ($s_{i}$ with $i = 1, 2, 3, 4$) together with the deformation
parameter, are introduced in the invariant. This allows for a larger variety
of interactions in the corresponding Hamiltonian problem.

\section{The basis states in the case of $q$-tensor $sp_{t}(4,R)$ algebra}

Now consider $\mathcal{H}$ as the space of the action of the $q$-deformed
tensor representation of the $sp_{t}(4,R)$ described in the previous
section. In terms of the spinor-like creation and annihilation operators, (
\ref{sp1}) and (\ref{sp2}), the basic states (\ref{cs}) have the form 
\begin{equation}
|\nu _{1},\nu _{-1}\rangle =q^{-\frac{1}{4}(\nu _{1}-\nu _{-1})-\frac{1}{2}
\nu _{1}\nu _{-1}}\frac{(t_{1}^{\dagger })^{\nu _{1}}(t_{-1}^{\dagger
})^{\nu _{-1}}}{\sqrt{[\nu _{1}]![\nu _{-1}]!}}|0\rangle ,  \label{ts}
\end{equation}
which introduces a dependence on $q$. It is easy to check that the operator $
P=(-1)^{N}$ commutes with all the generators, (\ref{a0}), (\ref{a1}) and (
\ref{a3}), so in the $q$-deformed tensor case the $sp_{t}(4,R)$
representation is also reducible and splits into two irreducible ones acting
in the $\mathcal{H}_{+}$ and $\mathcal{H}_{-}$ subspaces.

In what follows we will consider the space $\mathcal{H}_{+}$ to have $\nu
=\nu _{1}+\nu _{-1}$ $even$. The states of $\mathcal{H}_{-}$ can be obtained
from the ones in $\mathcal{H}_{+}$ with the help of the operators $
t_{k}^{\dagger }$ and $t_{k}$ ($k=\pm 1$). The later can be considered to be
the odd generators of the superalgebraic extension of the even $sp_{t}(4,R)$ 
\cite{7b}.

Looking forward to future applications, we represent the basic states $|\nu
_{1},\nu _{-1}\rangle \in \mathcal{H}_{+}$ as 
\begin{equation}
|n_{1,}n_{0},n_{-1}\rangle =\eta
(n_{1,}n_{0},n_{-1})(T_{1}^{1})^{n_{1}}(T_{0}^{1})^{n_{0}}(T_{-1}^{1})^{n_{-1}}|0\rangle \equiv |\nu _{1},\nu _{-1}\rangle
\label{spb}
\end{equation}
where \ n$_{i}$ ($i=0,\pm 1$) are integers restricted by the linkages 
\begin{equation}
\begin{tabular}{ll}
$\nu _{1}=2n_{1}+n_{0}$ & $\nu _{-1}=2n_{-1}+n_{0},$
\end{tabular}
\label{ml}
\end{equation}
and $\eta (n_{1,}n_{0},n_{-1})$ is the normalization factor given by 
\begin{equation}
\eta (n_{1,}n_{0},n_{-1})=\frac{q^{-\frac{1}{2}(n_{1}-n_{-1})-n_{1}\left(
n_{0}+n_{-1}\right) -n_{-1}\left( n_{0}+n_{1}\right) }}{\sqrt{
[2]^{n_{0}}[2n_{1}+n_{0}]![2n_{-1}+n_{0}]!}}.
\end{equation}
This representation of the basic states is useful for a consideration of
appropriate mapping procedures \cite{ggig} for the $sp_{t}(4,R)$ algebra. It
should be noted that in this case we use the ordering (\ref{spb}) of the
components of the vector $T_{m}^{1}$ ($m=0,\pm 1$) (\ref{a0}). Other
orderings can be obtained from this one by means of the commutation
relations (\ref{c17}). The $q$-factors that will appear in such a result can
be incorporated into the normalization coefficients.. Keeping in mind (\ref
{na} ) and (\ref{wcg}) it is obvious that 
\begin{equation}
\mathcal{N}_{1}|\nu _{1},\nu _{-1}\rangle =[\nu _{1}]q^{\nu _{-1}}|\nu
_{1},\nu _{-1}\rangle ;\mathstrut \;\mathcal{N}_{-1}|\nu _{1},\nu
_{-1}\rangle =[\nu _{-1}]q^{-\nu _{1}}|\nu _{1},\nu _{-1}\rangle  \label{nta}
\end{equation}
\begin{equation}
\lbrack N]|\nu _{1},\nu _{-1}\rangle =[\nu ]|\nu _{1},\nu _{-1}\rangle .
\label{ntat}
\end{equation}
Therefore, passing to representation (\ref{spb}) and in view of (\ref{ml} )
one finds 
\begin{eqnarray*}
\mathcal{N}_{1}|n_{1,}n_{0},n_{-1}\rangle
&=&[2n_{1}+n_{0}]q^{2n_{-1}+n_{0}}|n_{1,}n_{0},n_{-1}\rangle ;\; \\
\mathcal{N}_{-1}|n_{1,}n_{0},n_{-1}\rangle
&=&[2n_{-1}+n_{0}]q^{-2n_{1}-n_{0}}|n_{1,}n_{0},n_{-1}\rangle
\end{eqnarray*}
and 
\[
\lbrack N]|n_{1,}n_{0},n_{-1}\rangle =[2n]|n_{1,}n_{0},n_{-1}\rangle .
\]
Note that 
\[
\begin{tabular}{ll}
$2n=2n_{1}+2n_{0}+2n_{-1}=\nu $ & $j_{0}=\frac{1}{2}(\nu _{1}-\nu
_{-1})=n_{1}-n_{-1}$ .
\end{tabular}
\]

Now we can consider two extremal cases for possible values of the additional
quantum number $n_{0}.$

\begin{enumerate}
\item  $n_{0}$ takes on minimal values. Since we are in the space $\mathcal{H
}_{+}$ the integers $\nu _{1}$ and $\nu _{-1}$ are simultaneously even or
odd. For $\nu _{1}$ and$\ $ $\nu _{-1}$ $even$ ($\min n_{0}=0)$ and with the
help of (\ref{cb}) we obtain from (\ref{ts}) 
\begin{eqnarray}
|n_{1},0,n_{-1}\rangle &=&q^{-\frac{1}{2}(n_{1}-n_{-1})-2n_{1}n_{-1}}\frac{
(T_{1}^{1})^{n_{1}}(T_{-1}^{1})^{n_{-1}}}{\sqrt{[2n_{1}]![2n_{-1}]!}}
|0\rangle =  \label{smtte} \\
&&q^{-\frac{1}{4}(\nu _{1}-\nu _{-1})-\frac{1}{2}\nu _{1}\nu _{-1}}\frac{
(T_{1}^{1})^{\frac{\nu _{1}}{2}}(T_{-1}^{1})^{\frac{\nu _{-1}}{2}}}{\sqrt{
[\nu _{1}]![\nu _{-1}]!}}|0\rangle .  \nonumber
\end{eqnarray}
For $\nu _{1}$ and $\nu _{-1}-odd$ ($\min n_{0}=1)$ we find that 
\begin{eqnarray*}
|n_{1},1,n_{-1}\rangle &=&q^{-\frac{1}{2}(3n_{1}+n_{-1})-2n_{1}n_{-1}}\frac{
(T_{1}^{1})^{n_{1}}T_{0}^{1}(T_{-1}^{1})^{n_{-1}}}{\sqrt{
[2][2n_{1}+1]![2n_{-1}+1]!}}|0\rangle = \\
&&q^{-\frac{1}{4}(\nu _{1}-\nu _{-1}-2)-\frac{1}{2}\nu _{1}\nu _{-1}}\frac{
(T_{1}^{1})^{\frac{\nu _{1}-1}{2}}T_{0}^{1}(T_{-1}^{1})^{\frac{\nu _{-1}-1}{2
}}}{\sqrt{[2][\nu _{1}]![\nu _{-1}]!}}|0\rangle .
\end{eqnarray*}

In this way the $q$-deformed spinors are coupled to maximal degrees in $
n_{1} $ and $n_{-1}$ for the components $T_{1}^{1}$ and $T_{-1}^{1}$,
respectively. Representing the basis states in $\mathcal{H}_{+}$ as $
|n_{1},n_{0},n_{-1}\rangle $ vectors, in the case of $\min n_{0}=0$ or $1$
we can redraw the pyramid in Fig.1 in the following way\textbf{:} \vskip 
0.25cm \centerline{\bf Figure 3} 
\[
\begin{tabular}{|l|lll|l|lll|}
\hline
$\nu $/i$_{0}$ & 3 & \multicolumn{1}{|l}{2} & \multicolumn{1}{|l|}{1} & 0 & 
-1 & \multicolumn{1}{|l}{-2} & \multicolumn{1}{|l|}{-3} \\ \cline{2-8}
0 &  &  &  & $|0,0,0\rangle $ &  &  &  \\ \cline{4-6}
2 &  &  & \multicolumn{1}{|l|}{$|1,0,0\rangle $} & $|0,1,0\rangle $ & $
|0,0,1\rangle $ & \multicolumn{1}{|l}{} &  \\ \cline{3-7}
4 &  & \multicolumn{1}{|l}{$|2,0,0\rangle $} & \multicolumn{1}{|l|}{$
|1,1,0\rangle $} & $|1,0,1\rangle $ & $|0,1,1\rangle $ & \multicolumn{1}{|l}{
$|0,0,2\rangle $} & \multicolumn{1}{|l|}{} \\ \cline{2-8}\cline{2-8}
6 & $|3,0,0\rangle $ & \multicolumn{1}{|l}{$|2,1,0\rangle $} & 
\multicolumn{1}{|l|}{$|2,0,1\rangle $} & $|1,1,1\rangle $ & $|1,0,2\rangle $
& \multicolumn{1}{|l}{$|0,1,2\rangle $} & \multicolumn{1}{|l|}{$
|0,0,3\rangle $} \\ \cline{2-8}
$\vdots $ & $\vdots $ & \multicolumn{1}{|l}{$\vdots $} & 
\multicolumn{1}{|l|}{$\vdots $} & $\vdots $ & $\vdots $ & 
\multicolumn{1}{|l}{$\vdots $} & \multicolumn{1}{|l|}{$\vdots $}
\end{tabular}
\]
From the above illustration it is easy to see that we can obtain each state
from the left (right ) diagonals of the pyramid by the action on the minimal
weight state with the raising operators $T_{1}(T_{-1})$, respectively.

\item  $n_{0}$ takes on maximal values. In this case we have $n_{-1}=0$ or $
n_{1}=0$ at $\nu _{1}\neq \nu _{-1}$ and $n_{-1}=n_{1}=0$ at $\nu _{1}=\nu
_{-1}$. There are two possibilities:

On the left side of Fig.1, where $\nu _{1}\geq \nu _{-1}$ and the coupling
of $T_{0}^{1}(n_{-1}=0,\max n_{0}=\nu _{-1})$ is to the maximal degree, 
\begin{eqnarray}
|n_{1},n_{0},0\rangle &=&q^{-\frac{1}{2}n_{1}-n_{1}n_{0}}\frac{
(T_{1}^{1})^{n_{1}}(T_{0}^{1})^{n_{0}}}{\sqrt{
[2]^{n_{0}}[2n_{1}+n_{0}]![n_{0}]!}}|0\rangle =  \label{smt0} \\
&&q^{-\frac{1}{4}(\nu _{1}-\nu _{-1})-\frac{1}{2}(\nu _{1-}\nu _{-1})\nu
_{1}}\frac{(T_{1}^{1})^{\frac{1}{2}(\nu _{1}-\nu _{-1})}(T_{0}^{1})^{\nu
_{-1}}}{\sqrt{[\nu _{1}]![\nu _{-1}]![2]^{\nu _{-1}}}}|0\rangle .  \nonumber
\end{eqnarray}
For the states from the right from the central ladder ($\nu _{1}\leq \nu
_{-1},n_{1}=0,\max n_{0}=\nu _{1})$ we get the expressions
\end{enumerate}

\begin{eqnarray*}
|0,n_{0},n_{-1}\rangle &=&q^{-\frac{1}{2}n_{-1}-n_{-1}n_{0}}\frac{
(T_{0}^{1})^{n_{0}}(T_{-1}^{1})^{n_{-1}}}{\sqrt{
[2]^{n_{0}}[n_{0}]![2n_{-1}+n_{0}]!}}|0\rangle = \\
&&q^{-\frac{1}{4}(\nu _{1}-\nu _{-1})+\frac{1}{2}\nu _{1}(\nu _{1}-\nu
_{-1})}\frac{(T_{0}^{1})^{\nu _{1}}(T_{-1}^{1})^{\frac{1}{2}(\nu _{-1}-\nu
_{1})}}{\sqrt{[\nu _{1}]![\nu _{-1}]![2]^{\nu _{1}}}}|0\rangle .
\end{eqnarray*}
In this case ($\max n_{0}=\nu _{1}$or $\nu _{-1})$ the table of the basic
states has the form: \vskip 0.25cm \centerline{\bf Figure 4} 
\[
\begin{tabular}{|l|lll|l|lll|}
\hline
$\nu $/i$_{0}$ & 3 & \multicolumn{1}{|l}{2} & \multicolumn{1}{|l|}{1} & 0 & 
-1 & \multicolumn{1}{|l}{-2} & \multicolumn{1}{|l|}{-3} \\ \hline
0 &  &  &  & $|0,0,0\rangle $ &  &  &  \\ \cline{4-6}
2 &  &  & \multicolumn{1}{|l|}{$|1,0,0\rangle $} & $|0,1,0\rangle $ & $
|0,0,1\rangle $ & \multicolumn{1}{|l}{} &  \\ \cline{3-7}
4 &  & \multicolumn{1}{|l}{$|2,0,0\rangle $} & \multicolumn{1}{|l|}{$
|1,1,0\rangle $} & $|0,2,0\rangle $ & $|0,1,1\rangle $ & \multicolumn{1}{|l}{
$|0,0,2\rangle $} & \multicolumn{1}{|l|}{} \\ \hline
6 & $|3,0,0\rangle $ & \multicolumn{1}{|l}{$|2,1,0\rangle $} & 
\multicolumn{1}{|l|}{$|1,2,0\rangle $} & $|0,3,0\rangle $ & $|0,2,1\rangle $
& \multicolumn{1}{|l}{$|0,1,2\rangle $} & \multicolumn{1}{|l|}{$
|0,0,3\rangle $} \\ \hline
$\vdots $ & $\vdots $ & \multicolumn{1}{|l}{$\vdots $} & 
\multicolumn{1}{|l|}{$\vdots $} & $\vdots $ & $\vdots $ & 
\multicolumn{1}{|l}{$\vdots $} & \multicolumn{1}{|l|}{$\vdots $}
\end{tabular}
\]
This case corresponds to a coupling to maximal degree for the operator $
T_{0} $. With it we move along the columns by acting on the minimal weight
state an infinite number of times. These two forms for the basis states are
equivalent. The transition between Case 1 and Case 2 is realized by means of
the relation 
\begin{equation}
(T_{0}^{1})^{2}=q^{-1}[2]T_{1}^{1}T_{-1}^{1}.  \label{rtt}
\end{equation}

We now give the action of the $q$-deformed tensor representation of the
algebra $su_{t}(2)\sim so_{t}(3)$ with generators $L_{m}^{1}$ ($m=0,\pm 1)$.
First note that 
\[
\lbrack L_{m}^{1},N]=0,m=0,\pm 1.
\]
From decomposition (\ref{ren}) one can observe that in each subspace $
\mathcal{H}_{\nu }$ ($\nu =2n)$ of $\mathcal{H}_{+}$ an irreducible
representation of $su_{t}(2)$ acts. The eigenvalue of the second-order
Casimir operator for a given irreducible representation is $\frac{1}{[2]}[2n
][2n+2]$(\ref{l2}). Furthermore we know the action of the raising and
lowering operators $L_{\pm 1}^{1}$ for $n$ fixed in the case when $
n_{0}=\max n_{0}=\nu _{1}$ or $\nu _{-1}$. Along the rows given by $\nu =2n$
at $\nu _{1}\geq \nu _{-1}$ we move by acting with the operator ($
L_{-1}^{1})^{k}$ ($k\leq n$) on the highest weight state $
|n_{1}=n,0,0\rangle $: 
\[
(L_{-1}^{1})^{k}|n,0,0\rangle =q^{-\frac{1}{2}k(2n-k)}\sqrt{\frac{[2n]![k]!}{
[2n-k]!}}|n-k,k,0\rangle .
\]
Further, for $\nu _{1}\leq \nu _{-1}$: 
\[
(L_{-1}^{1})^{k}|0,n,0\rangle =q^{\frac{1}{2}k^{2}}\sqrt{\frac{[n+k]!}{[n-k]!
}}|0,n-k,k\rangle ,
\]
and therefore 
\[
(L_{-1}^{1})^{2n}|n,0,0\rangle =[2n]!|0,0,n\rangle .
\]
For the action of ($L_{+1}^{1})^{k}$ ($k\leq n$) on the lowest weight vector 
$|0,0,n\rangle $ at $\nu _{1}\leq \nu _{-1}$ we have that 
\[
(L_{+1}^{1})^{k}|0,0,n\rangle =q^{\frac{1}{2}k(2n-k)}\sqrt{\frac{[k]![2n]!}{
[2n-k]!}}|0,k,n-k\rangle ,
\]
and from the center for $\nu _{1}\geq \nu _{-1}$ we obtain the result 
\[
(L_{+1}^{1})^{k}|0,n,0\rangle =q^{-\frac{1}{2}k^{2}}\sqrt{\frac{[n+k]!}{
[n-k]!}}|k,n-k,0\rangle ,
\]
and it therefore follows that 
\[
(L_{+1})^{2n}|0,0,n\rangle =[2n]!|n,0,0\rangle .
\]
The operators $L_{\pm 1}^{1}$do not differ essentially from the operators $
J_{\pm }$ (\ref{rLc}) so their action on the basis states is easily obtained
by means of (\ref{ajpm}) if we take into account the appropriate $q$-factors
and the relations $j=\frac{\nu }{2}=n$ and $j_{0}=n_{1}-n_{-1}$. The
eigenvalues of the operator $L_{0}^{1}$ (\ref{l0r}) on the basis states are
given by 
\[
\begin{array}{c}
L_{0}^{1}|n_{1},n_{0},n_{-1}\rangle =\frac{1}{[2]}
\{q[2n_{1}+n_{0}][2n_{-1}+n_{0}+1]|n_{1},n_{0},n_{-1}\rangle \\ 
-q^{-1}[2n_{-1}+n_{0}][2n_{1}+n_{0}+1]|n_{1},n_{0},n_{-1}\rangle \} .
\end{array}
\]
Unlike $J_{0}$, $L_{0}^{1}$ has different eigenvalues for each step of a
given ladder.

\section{Conclusions}

In this paper the boson representation of the $sp(4,R)$ algebra and two
different deformations of it, $sp_{q}(4,R)$ and $sp_{t}(4,R)$, were
considered. The initial as well as the deformed representations act in the
same Fock space $\mathcal{H}$. All three are reducible and each one is
decomposed into two irreducible representations acting in the subspaces $
\mathcal{H}_{+}$ and $\mathcal{H} _{-}$ of $\mathcal{H}$.

The deformed representation $sp_{q}(4,R)$ is based on the standard $q$
-deformation of the two component Heisenberg algebra, realized in terms of
creation and annihilation operators. In this case eight of the ten
generators are deformed, but the complete set of the boson number operators $
N_{1}$ and $N_{-1}$ (or their linear combinations $N$ and $J_{0})$ are
preserved as in the ``classical'' case. The latter cannot be expressed in
terms of the deformed bosons. The subalgebras of the boson $sp(4,R)$ (the
compact $u(2)$ and the noncompact $u^{\varepsilon }(1,1)$ with $\varepsilon
=0,\pm $) are deformed as well and their deformed representations are
contained in $sp_{q}(4,R)$. They are reducible in the spaces $\mathcal{H}
_{+} $ and $\mathcal{H}_{-}$ and decompose into irreducible ones. In this
way a full description of the irreducible unitary representations of $
u_{q}(2)$ of the deformed ladder series $u_{q}^{0}(1,1)$ and of two deformed
discrete series $u_{q}^{\pm }(1,1)$ were obtained.

The other deformed representation, $sp_{t}(4,R)$, is realized by means of a
transformation of the $q$-deformed bosons into $q$-tensors (spinor-like)
with respect to $su_{q}(2)$ operators. Unlike $sp_{q}(4,R)$, the $
sp_{t}(4,R) $ generators are deformed and they have expressions in terms of
tensor products of spinor-like operators. The important result in this case
is the creation of a deformed $su_{t}(2)$, which can be interpreted as a
deformation of the angular momentum algebra $so(3)$. Its representation in $
\mathcal{H}$ is reducible and is decomposed into irreducible ones, giving in
this way their full description. In a future application, the dependence of
the two quantum number basis states in $\mathcal{H}_{+}$ will be presented
in terms of three linked integer parameters.

The reductions into subalgebras (compact and noncompact) of $sp(4,R)$ and
its deformations give rise to the possibility of different models with
dynamical symmetries. In any physical interpretations of the results it is
important to pay attention to the fact that the deformations do not change
the basis states in the Fock space, only the action of the operators on
them. This, with a view towards applications, gives rise to richer choices
for operators associated with observables - nondeformed, as well as
deformed. In a Hamiltonian theory this implies a dependance of matrix
elements on the deformation parameter, leading to the possibility of greater
flexibility and richer structures within the framework of algebraic
descriptions.

{\tiny {\bf Acknowledgments.} Supported in part by the U.S. National Science
Foundation through a regular grant, PHY-9970769, and a Cooperative Agreement,
EPS-9720652, that includes matching from the Louisiana Board of Regents Support
Fund. One of the authors /AIG/ acknowledges the useful discussions and help of V.
G. Gueorguiev.}

\end{document}